\pdfoutput=1
\documentclass[prb,twocolumn,showpacs,amsmath,amssymb,floatfix]{revtex4}

\usepackage{graphicx}%Include figure files
\usepackage{dcolumn}%Align table columns on decimal point
\usepackage{bm}% bold math
\usepackage{amsmath}
\usepackage{amssymb}

\begin{document}

%    \title{Finite Gilbert damping of perfect crystalline systems from ...: Application to ferromagnet/heavy-metal bilayers} 	
%	\title{Current Induced Damping Rate of Quantum Magnetic Moments in the Presence of Spin-Orbit Interaction}
 \title{Current Induced Damping of Nanosized Quantum Moments in the Presence of Spin-Orbit Interaction}
	
	\author{Farzad Mahfouzi}
	\email{Farzad.Mahfouzi@gmail.com}
	\affiliation{Department of Physics and Astronomy, California State University, Northridge, CA, USA}
	\author{Nicholas Kioussis}
	\email{nick.kioussis@csun.edu}
	\affiliation{Department of Physics and Astronomy, California State University, Northridge, CA, USA}
\date{\today}	
	\begin{abstract}
	
Motivated by the need to understand current-induced magnetization dynamics at the nanoscale, we have developed a formalism, within  the framework of Keldysh Green function approach, to study
the current-induced dynamics of a ferromagnetic (FM) nanoisland overlayer on a spin-orbit-coupling (SOC) Rashba plane. In contrast to the commonly employed classical micromagnetic LLG simulations
 the magnetic moments of the FM are treated {\it quantum mechanically}. We obtain the density matrix of the whole system consisting of conduction electrons entangled with the local magnetic moments and calculate the effective damping rate of the FM.
We investigate two opposite limiting regimes of FM dynamics: (1) The precessional regime where the magnetic anisotropy energy (MAE) and precessional frequency are smaller than the exchange interactions, and (2) The local spin-flip regime where the MAE and precessional frequency are comparable to the exchange interactions.
In the former case, we show that due to the finite size of the FM domain, the \textquotedblleft Gilbert damping\textquotedblright does not diverge in the ballistic electron transport regime, in sharp contrast to Kambersky's breathing Fermi surface theory for damping in metallic FMs. In the latter case, we show that above a critical bias the excited conduction electrons can switch the local spin moments resulting in demagnetization and reversal of the magnetization. Furthermore, our calculations show that the bias-induced antidamping efficiency in the local spin-flip regime is much higher than that in the rotational excitation regime.
	\end{abstract}

	\maketitle
	
	\section{Introduction}\label{sec:intro}
Understanding the current-induced magnetization switching (CIMS) at the nanoscale is mandatory for the scalability of non-volatile magnetic random access memory (MRAM) of the next-generation miniaturized spintronic devices. However, the local magnetic moments of a nanoisland require quantum mechanical treatment rather than the classical treatment of magnetization commonly employed in micromagnetic
simulations, which is the central theme of this work.

The first approach of CIMS employs the spin transfer torque (STT)~\cite{Slonczewski1996,Berger1996} in magnetic tunnel junctions (MTJ) consisting of two ferromagnetic (FM) layers (i.e., a switchable free layer and a fixed layer) separated by an insulating layer, which involves spin-angular-momentum transfer from conduction electrons to local magnetization~\cite{Ralph2008,Theodonis2006}. Although STT has proven very successful and brings the precious benefit of improved scalability, it requires high current densities ($\geq$ 10$^{10}$ A/cm$^2$) that are uncomfortably high for the MTJ's involved and hence high power consumption.
The second approach involves an in-plane current in a ferromagnet-heavy-metal bilayer where the magnetization switching is through the so-called spin-orbit torque (SOT) for both out-of-plane and in-plane magnetized layers.~\cite{Gambardella2011, Miron2010, Liu2012,Liu2012b} The most attractive feature of the SO-STT method is that the current does not flow through the tunnel barrier, thus offering potentially faster and more efficient magnetization switching compared to the MTJs counterparts.

As in the case of STT, the SO-STT has two components: a field-like and an antidamping component. While the field-like component reorients the equilibrium direction of the FM, the antidamping component provides the energy necessary for the FM dynamics by either enhancing or decreasing the damping rate of the FM depending on the direction of the current relative to the magnetization orientation as well as the structural asymmetry of the material. For sufficiently large bias the SOT can overcome the intrinsic damping of the FM leading to excitation of the magnetization precession.~\cite{Liu2012b} The underlying mechanism of the SOT for
both out-of-plane and in-plane magnetized layers remains elusive and is still under debate. It results from either the bulk Spin Hall Effect (SHE)~\cite{Dyakonov1971,Hirch1999,Jungwirth2012,Sinova2014}, or the interfacial Rashba-type spin-orbit coupling,~\cite{Bijl2012,Kim2012,Kurebayashi2014,MahfouziPRB2016} or both~\cite{Freimuth2014,Kim2013,Xiao2014}.
	
\begin{figure}
	\includegraphics[scale=0.3,angle=0]{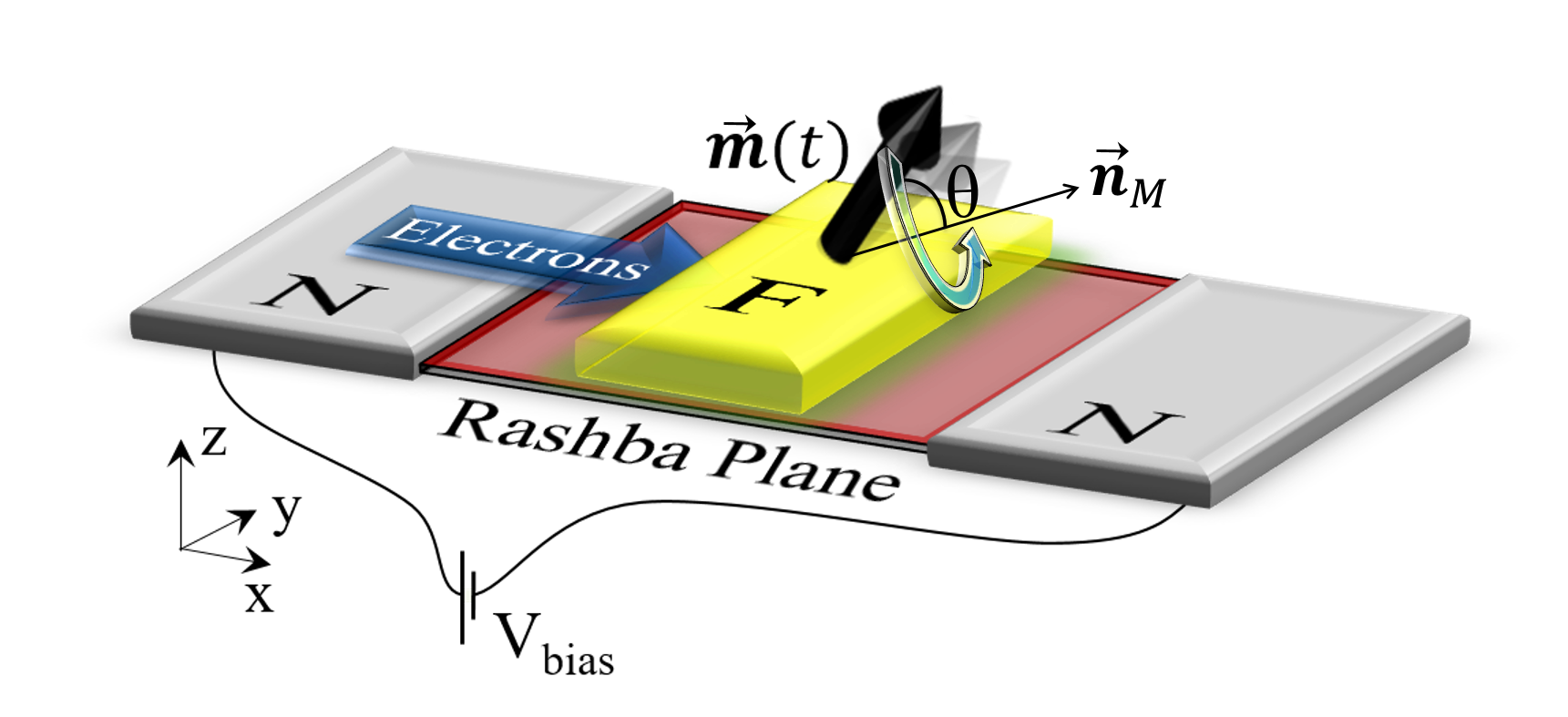}
		\caption{(Color online)
Schematic view of the FM/Rashba plane bilayer where the FM overlayer has length $L_x$ and is infinite (finite) along the $y$-direction for the case of a single domain (nano-island) discussed in Sec. \ref{sec:SingleDom}
(\ref{sec:LocMom}). The magnetization, $\vec{m}$, of the FM precesses around the direction denoted by the unit vector, $\vec{n}_M$,  with frequency $\omega$ and cone angle, $\theta$. The Rashba layer is attached
to two normal (N) leads which are semi-infinite along the $x$-direction, across which an external bias voltage, $V$, is applied.}
		\label{fig:fig1}
	\end{figure}

Motivated by the necessity of scaling down the size of magnetic bits and increasing the switching speed, the objective of this work is to develop a fully quantum mechanical formalism, based on the Keldysh Green function (GF) approach, to study the current-induced local moment
dynamics of a bilayer consisting of a FM overlayer on a SOC Rashba plane, shown in Fig. \ref{fig:fig1}.

Unlike the commonly used approaches to investigate the magnetization dynamics of quantum FMs, such as the master equation~\cite{Chudnovskiy2014}, the scattering~\cite{Fahnle2011} or quasi-classical~\cite{Swiebodzinski2010} methods, our formalism allows the study of magnetization dynamics in the presence of nonequilibrium flow of electrons.

%%%%%%%%%%%%%%%%%%%%%%%%%%%%%%%%%%%%%%%%%%%%%%%%  Old %%%%%%%%%%%%%%%%%%%%%%%%%%%%%%%%%%%%%%%

	%				%
	
We consider two different regimes of FM dynamics: In the first case, which we refer to as the single domain dynamics, the MAE and the precession frequency are smaller than the exchange interactions, and the FM can be described by a single quantum magnetic moment, of a typically large spin, $S$, whose dynamics are governed mainly by the quantized rotational modes of the magnetization. We show that the magnetic degrees of freedom entering the density matrix of the conduction electron-local moment entagled system simply shift the chemical potential of the Fermi-Dirac distribution function by the rotational excitations energies of the FM from its ground state. We also demonstrate that the effective damping rate is simply the {\it net} current along the the {\it auxiliary} $m$-direction, where $m$= -S, -S+1, $\ldots$, +S, are the eigenvalues of the total $S_z$ of the FM. Our results for the change of the damping rate due to the presence of a bias voltage are consistent with the anti-damping SOT of classical magnetic moments,~\cite{Li2015,MahfouziPRB2016}, where due to the Rashba spin momentum locking, the anti-damping SOT, to lowest order in magnetic exchange coupling, is of the form,  $\vec{m}\times(\vec{m}\times\hat{y})$, where $\hat{y}$ is an in-plane unit vector normal to the transport direction.

In the adiabatic and ballistic transport regimes due to the finite S value of the nanosize ferromagnet our formalism yields a finite \textquotedblleft Gilbert damping\textquotedblright, in sharp contrast to Kambersky's breathing Fermi surface theory for damping in metallic FMs.~\cite{Kambersky2007} On the other hand, Costa and Muniz ~\cite{Costa2015} and Edwards~\cite{Edwards2016}   demonstrated that the problem of divergent Gilbert damping is removed by taking into account the collective excitations. Furthermore, Edwards points out~\cite{Edwards2016}  the necessity of including the effect of long-range Coulomb interaction in calculating damping for large SOC.

In the second case, which corresponds to an independent local moment dynamics, the FM has a large MAE and hence the rotational excitation
energy is comparable to the local spin-flip excitation (exchange energy). We investigate the effect of bias on the damping rate of the local spin moments.
We show that above a critical bias voltage the flowing conduction electrons can excite (switch) the local spin moments resulting in demagnetization and reversal of the magnetization. Furthermore, we find
that, in sharp contrast to the single domain precessional dynamic, the current-induced damping is nonzero for in-plane and out-of-plane directions of
the equilibrium magnetization.
The bias-induced antidamping efficiency in the local moment switching regime is much higher than that in the single domain precessional dynamics.
	
The paper is organized as follows. In Sec.~\ref{sec:GenForm} we present the Keldysh formalism for the density matrix of the entagled quantum moment-conduction electron system and the effective dampin/antdamping torque.
In Sec.~\ref{sec:SingleDom} we present results for the current-induced damping rate in the single domain regime.  In Sec.~\ref{sec:LocMom} we present results for the current-induced damping rate in the independent
local regime.  We conclude in Sec.~\ref{sec:conclusions}.

\section{Theoretical Formalism}\label{sec:GenForm}
					
Fig. \ref{fig:fig1} shows a schematic view of the ferromagnetic heterostructure under investigation consisting of a 2D ferromagnet-Rashba plane bilayer attached to
two semi-infinite normal (N) leads whose chemical potentials are shifted by the external bias, $V_{bias}$. The magnetization of the FM precesses around the axis
specified by the unit vector, $\vec{n}_{M}$, with frequency $\omega$ and cone angle $\theta$. The FM has length, $L_x^{FM}$, along the transport direction.
The total Hamiltonian describing the coupled conduction electron-localized spin moment system in the heterostructure in Fig. \ref{fig:fig1}  can be written as,
				\begin{widetext}
				\begin{align}\label{eq:Eq1}
				 H_{tot}=\sum_{\bold{r}\bold{r}',\sigma\sigma'}Tr_{\{s_d\}}\left[\left(\boldsymbol{1}_s\hat{H}^{\sigma\sigma'}_{\bold{r}\bold{r}'}+\delta_{\bold{r}\bold{r}'}\delta_{\sigma\sigma'}\boldsymbol{1}_s\mu_{\bold{r}}+\delta_{\bold{r}\bold{r}'} J_{sd}\vec{\sigma}_{\sigma\sigma'}\cdot{\vec{\boldsymbol{s}}}_d(\bold{r})+\delta_{\sigma\sigma'}\delta_{\bold{r}\bold{r}'}\boldsymbol{H}_M\right){\psi}^*_{\{s_d'\}\bold{r}'\sigma'}{\psi}_{\{s_d\}\bold{r}\sigma}\right].
				\end{align}
				\end{widetext}	
Here, $\vec{\boldsymbol{s}}_d(\bold{r})$ is the local spin moment at atomic position $\bold{r}$,
 the trace is over the different configurations of the local spin moments, $\{s_d\}$,
${\psi}_{\{s_d\}\bold{r}\sigma}=|\{s_d\} \rangle\otimes\psi^e_{\bold{r}\sigma}$ is the quasi-particle wave-function associated with the conduction electron ($\psi^e$)
entangled to the FM states ($|\{s_d\} \rangle$), $ J_{sd}$ is the $s-d$ exchange interaction, $\boldsymbol{1}_s$ is the identity matrix in spin configuration space, and $\hat{\sigma}_{x,y,z}$ are the Pauli matrices. We use the convention that, except for $\bold{r}$, bold symbols represent operators in the magnetic configuration space and symbols with hat represent operators in the single particle Hilbert space of the conduction electrons. The magnetic Hamiltonian $\boldsymbol{H}_M$ is given by
\begin{align}\label{eq:MagEnrgy}
\boldsymbol{H}_M=&-g\mu_B\sum_{\bold{r}}\vec{B}^{ext}(\bold{r})\cdot{\vec{\boldsymbol{s}}}_d(\bold{r})\\
&-\sum_{\langle\bold{r},\bold{r}'\rangle}\frac{J^{dd}_{\bold{r}\bold{r}'}}{s_d^2}\vec{\boldsymbol{s}}_d(\bold{r}')\cdot\vec{\boldsymbol{s}}_d(\bold{r})-
\sum_{\bold{r}}\frac{J_{sd}}{s_d}\vec{\boldsymbol{s}}_c(\bold{r})\cdot\vec{\boldsymbol{s}}_d(\bold{r})\nonumber,
\end{align}
where, the first term is the Zeeman energy due to the external magnetic field, the second term is the magnetic coupling between the local moments and the third term is the energy associated
 with the intrinsic magnetic field acting on the local moment, $\vec{\boldsymbol{s}}_d(\bold{r})$, induced by the local spin of the conduction electrons, $\vec{\boldsymbol{s}}_c(\bold{r})$.

%%%%%%%%%%%%
 The Rashba model of a two-dimensional electron gas with spin orbit coupling interacting with a system of localized magnetic moments has been extensively employed~\cite{Haney2013,Park2013,Kim2012} to describe the effect of enhanced spin-orbit coupling solely at the interface on the current-induced torques in ultrathin ferromagnetic (FM)/heavy metal (HM) bilayers. The effects of (i) the ferromagnet inducing a moment in the HM and (ii) the HM with strong spin-orbit coupling inducing a large spin-orbit effect in the ferromagnet (Rashba spin-orbit coupling) lead to a thin layer where the magnetism and the spin-orbit coupling coexist.~\cite{Haney2013}
%%%%%%%%%%%%

The single-electron tight-binding Hamiltonian~\cite{Data_book} for the conduction electrons of the 2D Rashba plane, ${H}^{\sigma\sigma'}_{\bold{r}\bold{r}'}$ which is finite along the transport direction $x$ and infinite along the $y$ direction is of the form,
				\begin{align}
				&\hat{H}^{\sigma\sigma'}_{xx'}(k_ya)=[t\cos(k_ya)\delta_{\sigma\sigma'}-t_{so}\sin(k_ya){\sigma}^x_{\sigma\sigma'}]\delta_{xx'}\\
				&+t(\delta_{x,x'+1}+\delta_{x+1,x'})\delta_{\sigma\sigma'}+it_{so}(\delta_{x,x'+1}-\delta_{x+1,x'}){\sigma}^y_{\sigma\sigma'}.\nonumber
				\end{align}
Here, $x,x'$ denote atomic coordinates along the transport direction, $a$ is the in-plane lattice constant, and $t_{so}$ is the Rashba SOI strength. The values of the local effective exchange interaction, $J_{sd}$= 1eV,  and of the nearest-neighbor hopping matrix element, $t$=1 eV, represent a realistic choice for simulating the exchange interaction of 3$d$ ferromagnetic transition metals and their alloys (Fe, Co).~\cite{Zhang2004,Sanvito2005,Eastman1980} The Fermi energy, $E_F$=3.1 eV, is about 1 eV below the upper band edge at 4 eV consistent 
with the {\it ab initio} calculations of the (111) Pt surface\cite{Kokalj1999}.
Furthermore, we have used $t_{so}$=0.5 eV which yields a Rashba parameter, $\alpha_R$ = $t_{so}a \approx$1.4 eV\AA ~ ($a$=2.77 \AA~is the in-plane lattice constant of the (111) Pt surface)
consistent with the experimental value of about 1-1.5 eV\AA~\cite{Miron2011} and the {\it ab initio} value of 1 eV\AA~\cite{Park2013}.
However, because other experimental measurements for Pt/Co/Pt stacks report\cite{Haazen2013} a Rashba parameter which is an order of magnitude smaller, in Fig.3 we show the damping rate for different values of the Rashba SOI..
				For the results in Sec.~\ref{sec:LocMom}, we assume a real space tight binding for propagation along $y$-axis.
%%%%%%%%%%%%%%%%%%%%%%%%%%%%%%%%%%%%%%%%%%%%%%%%%%%%%%%%%%%%%%%%%%%%%%%%%%%%%%%%%%%%%%%%%%%%%%%%%%%%%%%%%%%%%%%%%%%%%				

The single particle propagator of the {\it coupled} electron-spin system is determined from the equation of motion of the retarded Green function,
				\begin{align}\label{eq:GFeq}
				&\left(E-i\eta-\hat{\mu}-\hat{H}-\boldsymbol{H}_M-\frac{{ J_{sd}}}{2}\hat{\vec{\sigma}}\cdot\hat{\vec{\boldsymbol{s}}}_{d}\right)\hat{\boldsymbol{G}}^r(E)=\hat{\boldsymbol{1}},
				\end{align}	
where, $\eta$ is the broadening of the conduction electron states due 
to inelastic scattering from defects and/or phonons, and for simplicity we ignore writing the identity matrices $\hat{1}$ and $\bold{1}$ in the expression.
				The density matrix of the entire system consisting of the noninteracting electrons (fermionic quasi-particles) and the local magnetic spins is determined (see
Appendix~\ref{AppA} for details of the derivation for a single FM domain) from the expression,
				\begin{align}\label{eq:GenDensMat}
				\hat{\boldsymbol{\rho}}=\int \frac{dE}{\pi}\hat{\boldsymbol{G}}^r(E)\eta {f}(E-\hat{\mu}-\boldsymbol{H}_M)\hat{\boldsymbol{G}}^a(E).
				\end{align}

It is important to emphasize that Eq. (\ref{eq:GenDensMat}) is the central result of this formalism which demonstrates that the effect of the local magnetic degrees of freedom is to shift the chemical potential of the Fermi-Dirac distribution function by the
eigenvalues, $\varepsilon_{m}$, of $\boldsymbol{H}_M|m\rangle=\varepsilon_{m}|m\rangle$, i.e.,
the excitation energies of the FM from its ground state. Here, $|m\rangle$ are the eigenstates of the Heisenberg model describing the FM.
The density matrix can then be used to calculate the local spin density operator of the conduction electrons, $[\vec{\boldsymbol{s}}_c(\bold{r})]^{mm'}=\sum_{ss'}\rho_{ss',\bold{r}\bold{r}}^{mm'}{\vec{\sigma}}_{ss'}/2$, which along with Eqs.~(\ref{eq:MagEnrgy}), ~(\ref{eq:GFeq}),
 and  ~(\ref{eq:GenDensMat}) form a closed set of equations that can be solved self consistently.
Since, the objective of this work is the damping/anti-damping (transitional) behavior of the FM in the presence of bias voltage, we only present results for the first iteration.

%%%%%%%%%%%%%%%%%%%%%%%%%%%%%%%%%%%%%%%%%%%%%%%%%%%%%%%%%%%%%%%%%%%%%%%%%%%%%%%%%%%%%%%%%%%%%%%%%%%%%%%%		
Eq.~(\ref{eq:GenDensMat}) shows that the underlying mechanism of the damping phenomenon is the flow of conduction electrons from states of
 higher chemical potential to those of lower one where the FM state relaxes to its ground state by transferring energy to the conduction electrons. Therefore, the FM dynamical properties in this formalism is completely governed by its coupling to the conduction electrons, where conservation of energy and angular momentum dictates the excitations as well as the fluctuations of the FM sate through the Fermi distribution function of the electrons coupled to the reservoirs. This is different from the conventional Boltzmann distribution function which is commonly used to investigate the thermal and quantum fluctuations of the magnetization.

 Due to the fact that the number of magnetic configurations (i.e. size of the FM Hilbert space) grows exponentially with the dimension of the system it becomes prohibitively expensive to consider all
 possible eigenstates of the $\boldsymbol{H}_M$ operator.  Thus, in the following sections we consider two opposite limiting cases of magnetic configurations. In the first case we assume a single magnetic moment for the whole FM which is valid for small FMs with strong exchange coupling between local moments and small MAE. In this case the dynamics is mainly governed by the FM rotational modes and local spin flips can be ignored. In the second case we ignore the correlation between different local moments and employ a mean field approximation such that at each step we focus on an individual atom by considering the local moment under consideration as a quantum mechanical object while the rest of the moments are treated classically. We should mention that a more accurate modeling of the system should contain both single domain rotation of the FM as well as the local spin flipping but also the effect of nonlocal correlations between the local moments and conduction electrons, which are ignored in this work.

				\section{Single Domain Rotational Switching}\label{sec:SingleDom}
				\begin{figure}
					\includegraphics[scale=0.4,angle=0]{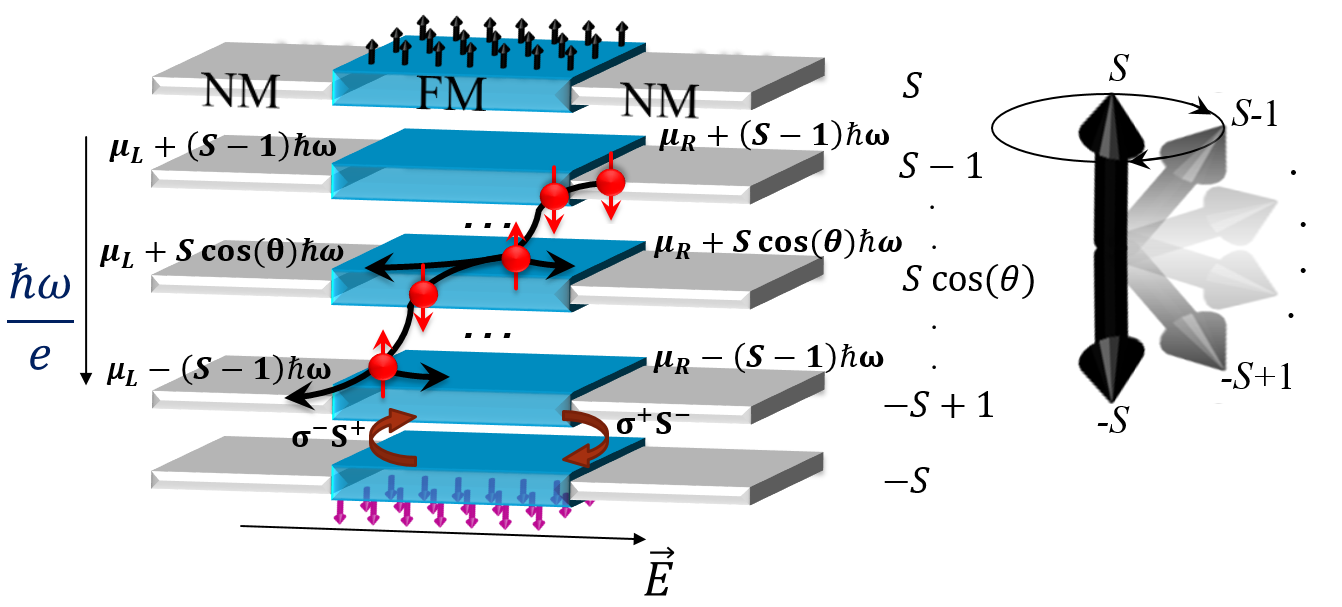}
					\caption{(Color online) Schematic representation of the quasiparticles of the FM and conduction electron entangled states. The horizontal planes denote the eigenstates, $|S,m\rangle$
of the total $\boldsymbol{S}_z$ of the FM with eigenvalues $m = -S, -S+1,\ldots ,+S$ along the {\it auxiliary} $m$-direction.
Excitation of magnetic state induces a shift of the chemical potential of the Fermi-Dirac distribution function leading to flow of quisiparticles along the $m$-direction
which corresponds to the damping rate of the FM. The FM damping involves two processes: (1) An intra-plane process involving spin reversal of the conduction electron via
the SOC; and (2) An inter-plane process involving quasiparticle flow of majority (minority) spin along the ascending (descending) $m$-direction due to
conservation of total angular momentum, where the interlayer hopping is accompanied by a spin flip of conduction electrons.}
					\label{fig:fig2}
				\end{figure}

In the regime where the energy required for the excitation of a single local spin moment ($\approx meV$) is much larger than the MAE ($\approx \mu eV$) the
low-energy excited states correspond to rotation of the {\it total} angular momentum of the FM acting as a single domain and
the effects of local spin flips described as the second term in Eq~\ref{eq:MagEnrgy}, can be ignored. In this regime all of the local moments behave collectively and the local moment operators can be replaced by the average spin operator, $\vec{\boldsymbol{s}}_d(\bold{r})=\sum_{\bold{r}'}\vec{\boldsymbol{s}}_d(\bold{r}')/N_{d}=s_d\vec{\boldsymbol{S}}/S$, where $N_{d}$ is the number of local moments and $\vec{\boldsymbol{S}}$ is the total angular momentum with amplitude $S$. The magnetic energy operator is given by $\boldsymbol{H}_M=-\vec{B}\cdot\boldsymbol{S}$, where, $\vec{B}=g\mu_B\vec{B}^{ext}+J_{sd}\vec{s}_c$. Here, for simplicity we assume $\vec{s}_c$ to be scalar and independent of the FM state. The eigenstates of $\boldsymbol{H}_M$ operator are then simply the
eigenstates, $|S,m\rangle$, of the total angular momentum $\boldsymbol{S}_z$, with eigenvalues $m\omega = -S\omega,\ldots ,+S\omega$, where $\omega=B_z$ is the Larmor frequency. Thus, the wave function of the coupled
electron-spin configuration system, shown schematically in Fig.~\ref{fig:fig2} is of the form, ${\psi}_{ms'\bold{r}}(t)=|S,m\rangle\otimes \psi_{s'\bold{r}}(t)$. One can see
that the magnetic degrees of freedom corresponding to the different eigenstates of the $\boldsymbol{S}_z$ operator,
 enters as an additional auxiliary dimension for the electronic system where the variation of the magnetic energy, $\langle S,m|\boldsymbol{H}_M|S,m\rangle=m\omega$, shifts
 the chemical potentials of the electrons along this dimension. The gradient of the chemical potential along the
 auxiliary direction, is the Larmor frequency ($\mu eV\approx GHz$) which appears as an effective \textquotedblleft electric field\textquotedblright  in that direction.

Substituting Eq~(\ref{eq:GenDensMat}) in Eq~\eqref{eq:trans}(b) and averaging over one precession period we find that the
 average rate of angular momentum loss/gain, which we refer to as the effective \textquotedblleft{\it damping rate}\textquotedblright per magnetic moment, can be written as

				\begin{align}
				\mathcal{T}_m&=\frac{1}{2}\Im(\mathcal{T}^-_{m}-\mathcal{T}^+_{m}),
                \label{eq:DampRate}
                \end{align}
where,
                \begin{align}
				\mathcal{T}^{\pm}_{m}&=\frac{J_{sd}}{2SN_d}Tr_{el}[ \hat{\sigma}^{\mp}{\boldsymbol{S}}^{\pm}_m\hat{\boldsymbol{\rho}}_{m,m\pm 1}].
				%\mathcal{J}_{+,m}&=Tr_{el}[{ J_{sd}}\hat{\sigma}^{-}{\boldsymbol{S}}_{m-1,m}^+\hat{\boldsymbol{\rho}}_{mm-1}(t)]\nonumber,
				\end{align}
is the current along the {\it auxiliary} $m$-direction in Fig.~\ref{fig:fig2} from the $m\leftrightarrow m+1$ ($\pm$ sign) state of the total $\boldsymbol{S}_z$ of the FM.
Here, $Tr_{el}$, is the trace over the conduction electron degrees of freedom, and ${\boldsymbol{S}}_m^{\pm}=\sqrt{S(S+1)-m(m\pm1)}$ are the ladder operators.
It is important to note that within this formalism the damping rate is simply the {\it net} current across the $m^{th}$-layer along the auxiliary direction associated with the transition rate of the FM from state $m$ to its nearest-neighbor states ($m\pm 1$).
				
\begin{figure}
					\includegraphics[scale=0.4,angle=0]{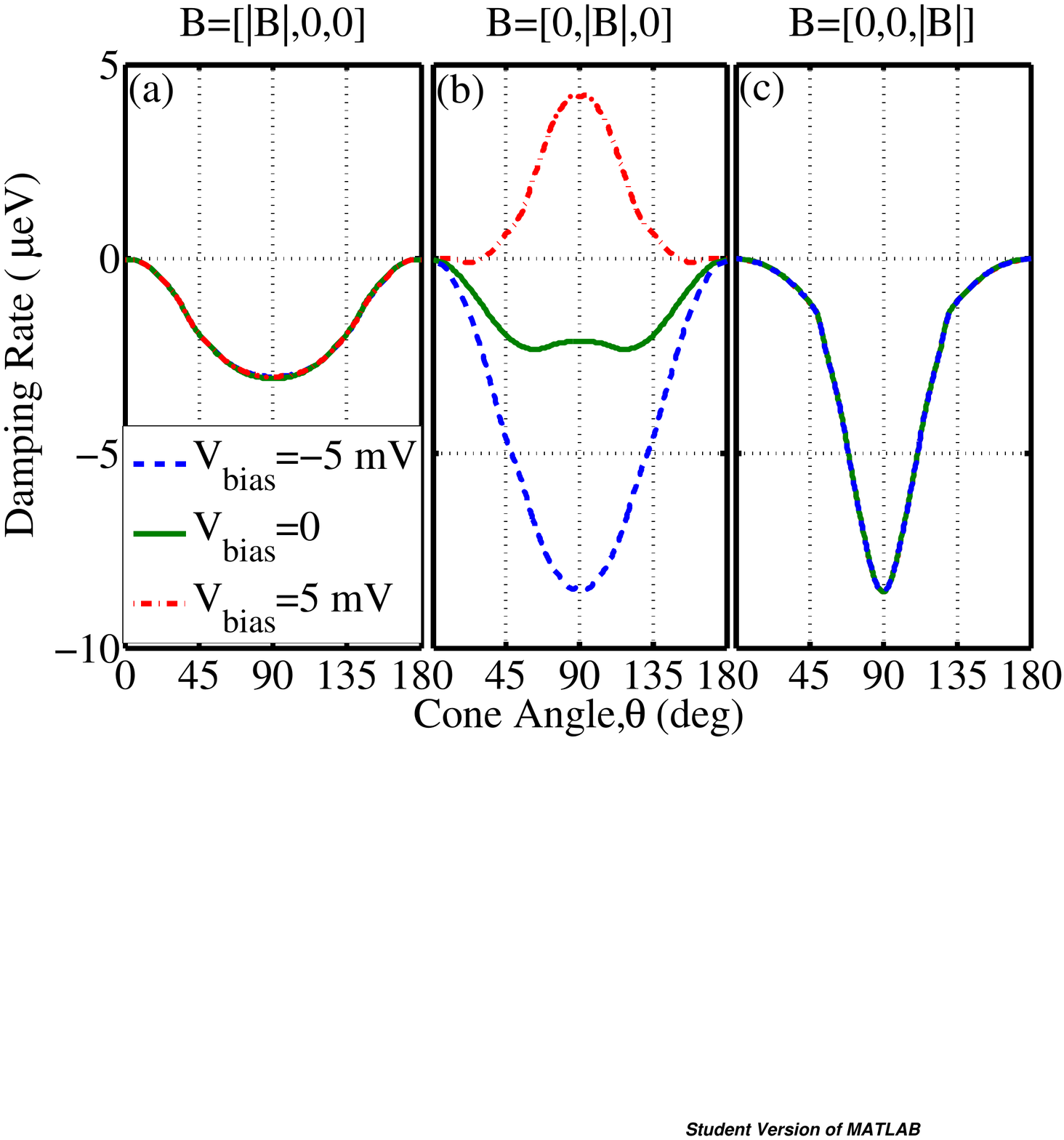}
					\caption{(Color online) Effective damping rate for a single FM domain
as a function of the precession cone angle, $\theta$, for various bias values
under an effective magnetic field which is in-plane (a) along and (b) normal to the transport direction and (c) out-of-plane. The length of the FM along the $x$ direction is $L_x=25 a$ while it is assumed to be infinite in the $y$-direction, $\hbar\omega=10\mu eV$, the broadening
parameter $\eta=0$, $k_BT=10 meV$ and the domain magnetic moment $S=200$. The results are robust with larger values of $S$ in either the ballistic, $\eta\ll\hbar\omega$, or dirty, $\eta\gg\hbar\omega$, regimes.}
\label{fig:fig3}
				\end{figure}

%%%%%%%%%%%%%%%%%%%%%%%%%%%%%%%%%%%%%%%%%%%%%%%%%%%%%%%%%%%%%%%%%%%%%%%%%%%%%%%%      FIG	
Fig. \ref{fig:fig3} shows the damping rate as a function of the precession cone angle, $\theta$ = cos$^{-1}(\frac{m}{S})$, for different values of bias and for
an in-plane effective magnetic field (a) along and (b) normal to the transport direction, and (c) an out-of-plane magnetic field.
For cases (a) and (c) the  damping rate is negative and relatively independent of bias for low bias values. A negative damping rate implies that the FM relaxes towards the magnetic field by losing its angular momentum, similar to the Gilbert damping rate term in the classical LLG equation, where its average value over the azimuthal precession angle, $\varphi=\omega t$, is of the form,
$\mathcal{T}=-\alpha s_d\int \frac{d\varphi}{2\pi} \vec{m}\times(\vec{m}\times \vec{B})\cdot\vec{n}_{M}$, which is nonzero (zero) when the unit vector $\vec{n}_{M}$ is along (perpendicular to) the effective magnetic field.
The dependence of the damping rate on the bias voltage when the effective magnetic field ${\vec{B}}$ is inplane and normal to the transport direction can be understood by the spin-flip reflection mechanism accompanied by Rashba spin-momentum locking described in Ref.~\cite{MahfouziPRB2016}. One can see that a large enough bias can result in a sign reversal of the damping rate and hence a magnetization reversal of the FM.
%In all cases $\mathcal{T} \propto\sin^2(\theta)$ similar to the Gilbert damping and the Slonczewski torques in LLG equation.
It's worth mentioning that due to the zero-point quantum fluctuations of the magnetization, at $\theta=0,\pi$ (i.e. $m=\pm S$) we have $\mathcal{T}\neq$0 which is inversely proportional to the size of the magnetic moment, $S$.

%%%%%%%%%%%%%%%%%%%%%%%%%%%%
%For simplicity and also to reduce the possible finite transverse size effect of the electronic wavefunction, we consider the plane to be infinite along the transverse direction. For the longitudinal length of the FM in the numerical calculation we choose $d_x=10a$ where $a$ is the lattice constant.
%%%%%%%%%%%%%%%%%%%%%%%%%%%%%%%%%%%%%%%%%%%%%%%%%%%%%%%%%%%%%%%%%%%%%%%

\begin{figure}
					\includegraphics[scale=0.4,angle=0]{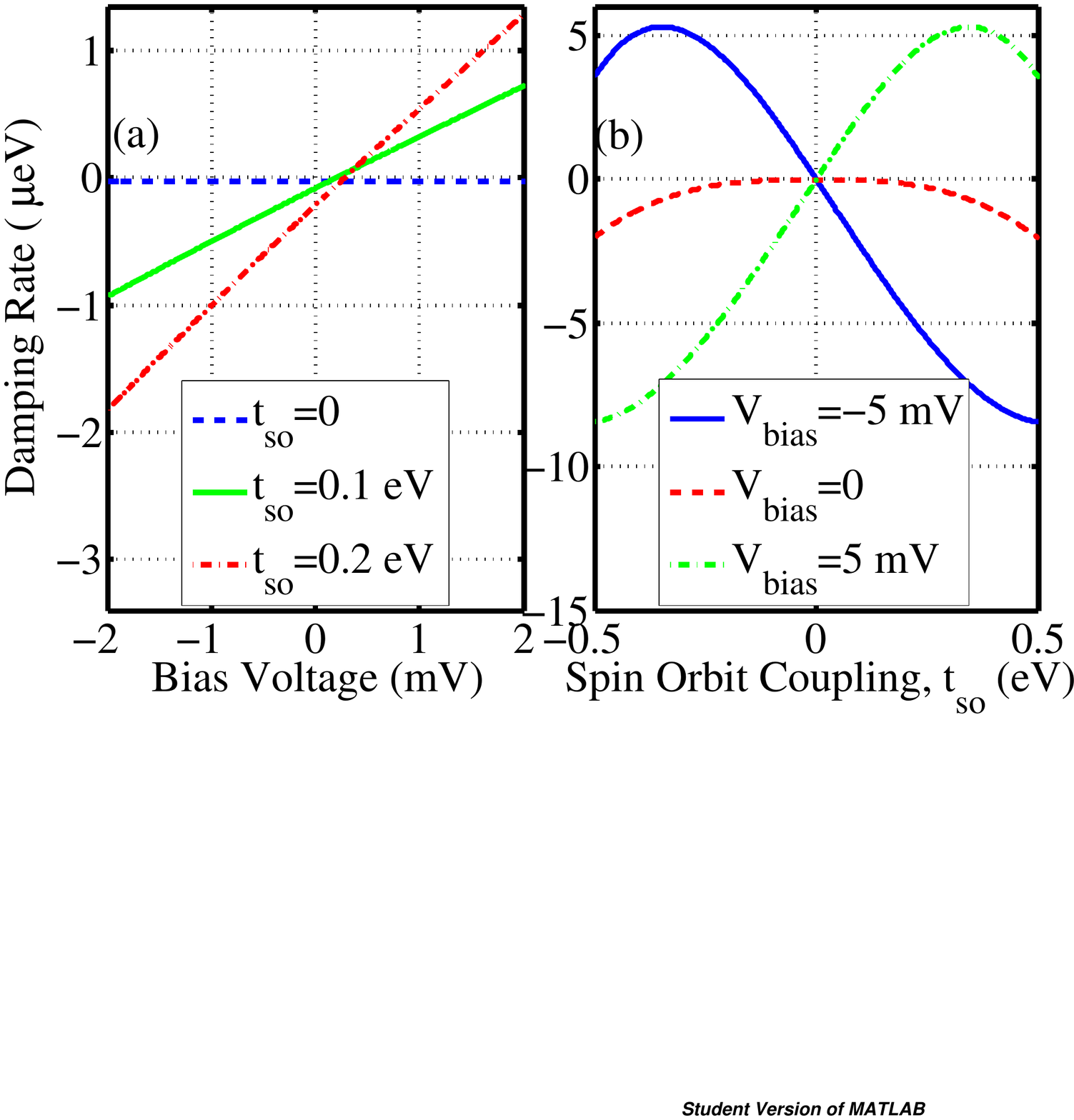}
					\caption{(Color online) Damping torque versus (a) bias voltage and (b) spin-orbit coupling strength, for $m=0$  corresponding to the precession cone angle of $90^o$. The precession axis of the FM is along the $y$-direction and the rest of the parameters are the same as in Fig.~\ref{fig:fig3}. The zero-bias damping rate versus SOC shows a $t_{so}^2$ dependence while the damping rate under non-zero bias exhibits nearly linear SOC dependence.}
					\label{fig:fig4}
				\end{figure}

%%%%%%%%%%%%%%%%%%%%%%%%%%%%%%%%%%%%%%%%%%%%%%%%%%%%%%%%%%%%%%%%%%%%%%%%%%%%%%%%%%  Figure 3 %%%%%%%%%%%%%%%%%%%%%%%%%%%%%%%%%%%%%%		
	In Fig. \ref{fig:fig4}(a) we present the {\it effective} damping rate versus bias for different values of the Rashba SOC. The results show a linear response regime with respect to the bias voltage where both the zero-bias damping rate and the slope, $d\mathcal{T}/dV$ increases with the Rashba SOC. This is consistent with Kambersky's mechanism of Gilbert damping due to the SOC of itinerant electrons,~\cite{Kambersky2007} and the SOT mechanism~\cite{MahfouziPRB2016}. Fig. \ref{fig:fig4}(b) shows that in the absence of bias voltage the damping rate is proportional to $t_{so}^2$ and the effect of the spin current pumped into the left and right reservoirs is negligible. This result of the $t_{so}^2$ dependence of the zero-bias damping rate is in agreement with recent calculations of Costa and Muniz\cite{Costa2015} and Edwards\cite{Edwards2016} which
took into account the collective excitations.
% and where the later\cite{Edwards2016} point out the necessity of including the effect of long-range Coulomb interactions.
In the presence of an external bias, $\mathcal{T}$ varies linearly with the SOC, suggesting that to the lowest order it can be fitted to
\begin{equation}\label{eq:fit}
\mathcal{T}=\sin^2(\theta) t_{so}(c_1 t_{so}\hbar\omega+c_2eV_{bias}),
\end{equation}	
where $c_1$ and $c_2$ are fitting parameters.		
%%%%%%%%%%%%%%%%%%%%%%%%%%%%%%%%%%%%%%%%%%%%%%%%%%%%%%%%%%%%%%%%%%%%%%%%%%

\begin{figure}
					\includegraphics[scale=0.4,angle=0]{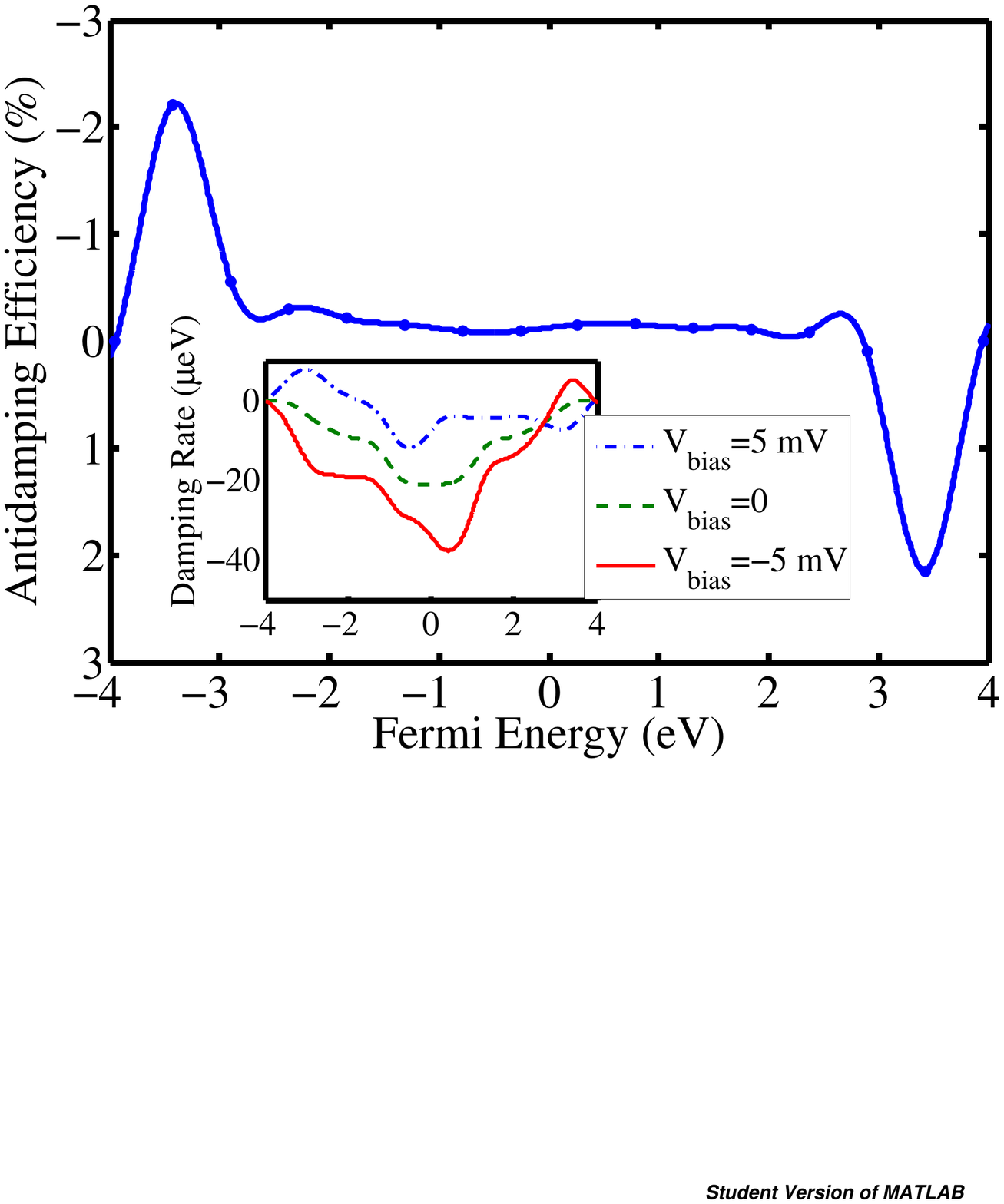}
					\caption{(Color online) Bias-induced precessional anti-damping efficiency, $\Theta=\hbar\omega(\mathcal{T}(V_{bias})-\mathcal{T}(0))/eV_{bias}\mathcal{T}(0)$, versus the Fermi energy of the 2D Rashba plane in Fig.\ref{fig:fig1}, where the energy band ranges from -4 eV to +4 eV. The magnetization precesses around the in-plane direction ($y-$axis) normal to the transport direction  and the rest of the parameters of the system are the same as in Fig.~\ref{fig:fig3}. Note, for magnetization precession around the $x$ and $z$ axis, $\mathcal{T}(V_{bias})=\mathcal{T}(0)$ for all precession cone angles and hence $\Theta$=0. Inset shows the damping rate versus the Fermi energy for different bias values used to calculate precessional anti-damping efficiency.}
					\label{fig:fig5}
				\end{figure}

The bias-induced efficiency of the anti-damping SOT,  $\Theta\equiv\hbar\omega(\mathcal{T}(V_{bias})-\mathcal{T}(0))/eV_{bias}\mathcal{T}(0)$,
describes how efficient is the energy conversion between the magnetization dynamics and the conduction electrons. Accordingly, for a given bias-induced efficiency, $\Theta$, one needs to apply an external bias equal to $\hbar\omega/e\Theta$  to overcome the zero-bias damping of the FM. Fig. \ref{fig:fig5} displays the anti-damping efficiency versus the position of the Fermi energy of the FM from the bottom (-4$t$=-4 eV) to the top (4$t$=4eV) of the conduction electron band for the two-dimensional square lattice. The result is independent of the bias voltage and the Larmor frequency in the linear response regime ({\it i.e.} $V_{bias},\omega\ll t$). We find that the efficiency peaks when the Fermi level is in the vicinity of the bottom or top of the energy band where the transport is driven by electron- or hole-like carriers and the Gilbert damping is minimum. The sign reversal of the antidamping SOT is due to the electron- or hole-like driven transport similar to the Hall effect.~\cite{Kittel_book}

	\textit{Classical Regime of the Zero Bias Damping rate ---}			
		In the following we show that in the case of classical magnetic moments ($S\rightarrow \infty$) and the adiabatic regime ($\omega\rightarrow 0$), the formalism developed in this paper leads to the conventional expressions for the damping rate. In this limit the system becomes locally periodic and one can carry out a Fourier transformation from $m\equiv S_z$ space to azimuthal angle of the magnetization orientation, $\varphi$, space. Conservation of the angular momentum suggests that the majority- (minority-) spin electrons can propagate only along the ascending (descending) $m$-direction, where the hopping between two nearest-neighbor $m$-layers is accompanied by a spin-flip. As shown in Fig. ~\ref{fig:fig2} the existence of spin-flip hopping requires the presence of intralayer SOC-induced noncollinear spin terms which rotate the spin direction of the conduction electrons as they propagate in each $m$-layer. This is necessary for the persistent flow of electrons along the $\varphi$ auxiliary direction and therefore damping of the magnetization dynamics. Using the Drude expression of the longitudinal conductivity along the $\varphi$-direction for the damping rate, we find that,
within the relaxation time approximation, $\eta/\omega\rightarrow\infty$, where the relaxation time of the excited conduction electrons is much shorter than the time scale of the FM dynamics,
$\mathcal{T}$ is given by
			\begin{equation}\label{eq:Classical_Damp}
				\mathcal{T}=-\frac{\omega}{\eta}\sum_{n}\int\frac{dk_xdk_yd\varphi}{(2\pi)^3}(v^{\varphi}_{n\vec{k}})^2f'(\varepsilon_{n\vec{k}}(\varphi)).
				\end{equation}	
		Here, $v_{n\vec{k}}^{\varphi}=\partial\varepsilon_{n\vec{k}}(\varphi)/\partial \varphi$  is the group velocity along the $\varphi$-direction in Fig. \ref{fig:fig2}, and $\varepsilon_{n,\vec{k}}=\varepsilon^0(|\vec{k}|)\pm|\vec{h}(\vec{k})|$ for the 2D-Rashba plane,
where $\varepsilon^0(|\vec{k}|)$ is the spin independent dispersion of the conduction electrons
and $\vec{h}=at_{so}\hat{e}_z\times\vec{k}+\frac{1}{2} J_{sd}\vec{m}$, is the spin texture of the electrons due to the SOC and the $s-d$ exchange interaction.
For small precession cone angle, $\theta$, the Gilbert damping constant can be determined from $\alpha=-\mathcal{T}/s_d\omega\sin^2(\theta)$, where the zero-temperature $\mathcal{T}$ is evaluated by Eq. (\ref{eq:Classical_Damp}).
We find that
				 \begin{equation}\label{eq:Gil_Damp}
				 \alpha\approx\frac{1}{\eta} t_{so}^2\left[(k^+_{F}a)^2D^+(E_F)+(k^-_{F}a)^2D^-(E_F)\right](1+\cos^2(\gamma)),
				 \end{equation}				
				 where $D^{+(-)}(E)$ is the density of states of the majority (minority) band, $\gamma$ is the angle between the precession axis and the normal to the Rashba plane, and the Fermi wave-vectors ($k^{\pm}_F$) are obtained from, $\varepsilon_0(k^{\pm}_F)=E_F\mp J_{sd}/2$. Eq.~(\ref{eq:Gil_Damp}) shows that the Gilbert damping increases as the precession axis changes from in-plane ($\gamma=\pi/2$) to out of plane ($\gamma=0$),~\cite{mahfouzi_SPIN2016} which can also be seen in Fig.~\ref{fig:fig3}.

			\begin{figure}
					\includegraphics[scale=0.4,angle=0]{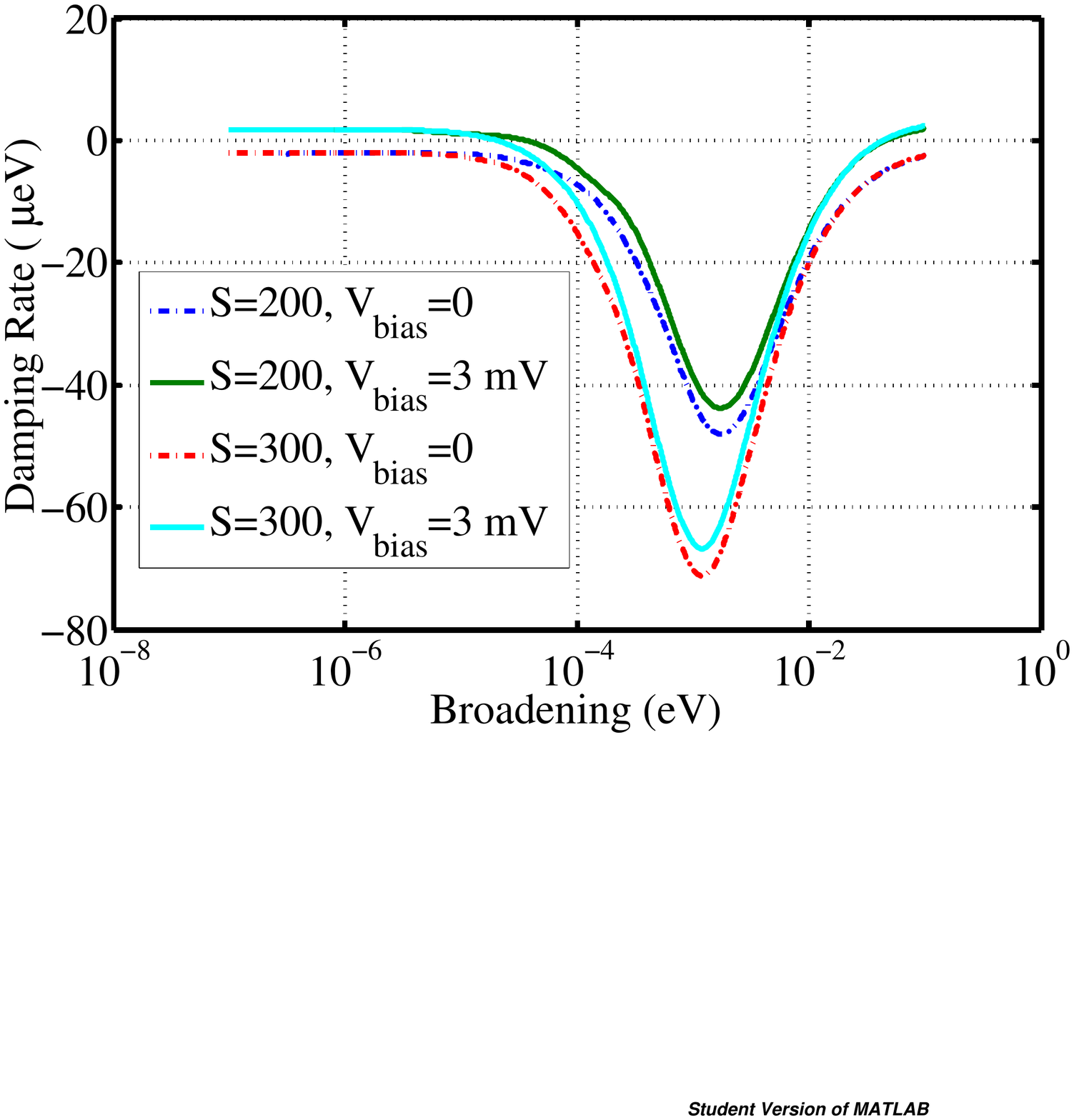}
					\caption{(Color online) Precessional damping rate versus broadening of the states in the presence (solid lines) and absence (dashed lines) of bias voltage for two values of the domain size $S=200$ and $S=300$. In both ballistic, $\eta/\omega\approx 0$, and diffusive, $\eta/\omega\gg 1$,  regimes the precessional damping rate is independent of the domain size, while in the intermediate case, the amplitude of the minimum of damping rate shows a linear dependence versus $S$. Note that the value of the broadening at which the damping rate is minimum varies inversely proportional to the domain size, $S$.}
					\label{fig:fig6}
				\end{figure}

It is important to emphasize that in contrast to Eq.~(\ref{eq:Classical_Damp}) the results shown in Fig.~\ref{fig:fig4} correspond to the ballistic regime with $\eta=0$  in the central region and the relaxation of the excited electrons occurs solely inside the metallic reservoirs. To clarify how the damping rate changes from the ballistic to the diffusive regime we present in Fig. \ref{fig:fig6} the damping rate versus the broadening, $\eta$, of states in the presence (solid line) and absence (dashed line) of bias voltage. We find that in both ballistic ($\eta/\omega\approx 0$) and diffusive ($\eta/\omega\gg 1$) regimes the damping rate is independent of the size of the FM domain, $S$. On the other hand, in the intermediate regime the FM dynamics become strongly dependent on the effective domain size where the minimum of the damping rate varies linearly with $S$.
This can be understood by the fact that the effective chemical potential difference between the first, $m=-S$ and last, $m=S$ layers in Fig.\ref{fig:fig3} is proportional to $S$ and for a coherent electron transport the conductance is independent of the length of the system along the transport direction. Therefore, in this case the FM motion is driven by a {\it coherent} dynamics.

				\section{Demagnetization Mechanism of Switching}\label{sec:LocMom}
In Sec. \ref{sec:SingleDom} we considered the case of a single FM domain where its low-energy excitations, involving
the precession of the {\it total} angular momentum, can be described  by the eigenstates $|m\rangle$ of $\boldsymbol{S}_z$ and local
spin flip processes were neglected.  However, for ultrathin FM films or FM nanoclusters, where the MAE per atom ($\approx meV$) is comparable
to the exchange energy between the local moments (Curie temperature), the low-energy excitations involve both magnetization rotation
and local moments spin-flips due to conduction electron scattering which can in turn change also $S$. In this case the switching is accompanied by the excitation of {\it local} collective modes that effectively lowers the amplitude of the magnetic ordering parameter.
For simplicity we employ the mean field approximation for the 2D FM nanocluster where the spin under consideration at position $\bold{r}$ is treated quantum mechanically
interacting with all remaining spins through an effective magnetic field, $\vec{B}$.
The spatial matrix elements of the local spin operator are
\begin{align}\label{eq:s_d}
&[\hat{\vec{\boldsymbol{s}}}_{d,\bold{r}}]_{\bold{r}_1\bold{r}_2}=\vec{s}_d(\bold{r}_1)\delta_{\bold{r}_1\bold{r}_2}(1-\delta_{\bold{r}_1\bold{r}})
\boldsymbol{1}_s+\frac{1}{2}\delta_{\bold{r}_1\bold{r}_2}\delta_{\bold{r}_1\bold{r}}\boldsymbol{\vec{\tau}},
\end{align}		
where, $\vec{\boldsymbol{\tau}}$s are the Pauli matrices. The magnetic energy can be expressed as, $\boldsymbol{H}_M(\bold{r})=-\vec{B}(\bold{r})\cdot\vec{\boldsymbol{\tau}}/2$,
where, the effective local magnetic field is given by,
\begin{align} \label{eq:B(r)}
\vec{B}(\bold{r})=g\mu_B\vec{B}^{ext}+4\sum_{\bold{r}'}J^{dd}_{\bold{r}\bold{r}'}\vec{s}_d(\bold{r}')+2J_{sd}\vec{s}_c(\bold{r}).
\end{align}		

The equation of motion for the single particle propagator of the electronic wavefunction entangled with the local spin moment under consideration can then be obtained from,
				\begin{align}
				&\left( E-\hat{\mu}-\boldsymbol{H}_{M}(\bold{r})-\hat{H}-\frac{{ J_{sd}}}{2}\hat{\vec{\sigma}}\cdot\hat{\vec{\boldsymbol{s}}}_{d,\bold{r}}\right)\hat{\boldsymbol{G}}^r_{\bold{r}}(E)=\hat{\boldsymbol{1}}.
				\end{align}		

	The density matrix is determined from Eq.~(\ref{eq:GenDensMat}) which can in turn be used to calculate the spin density of the conduction electrons, $\vec{s}_c(\bold{r})=Tr(\hat{\vec{\sigma}}\hat{\boldsymbol{\rho}}_{\bold{r}\bold{r}})/2$, and the direction and amplitude of the local magnetic moments, $\vec{s}_d(\bold{r})=Tr(\vec{\boldsymbol{\tau}}\hat{\boldsymbol{\rho}}_{\bold{r}\bold{r}})/2$. 		

				\begin{figure}
					\includegraphics[trim={0 0 0 0},scale=0.4,angle=0]{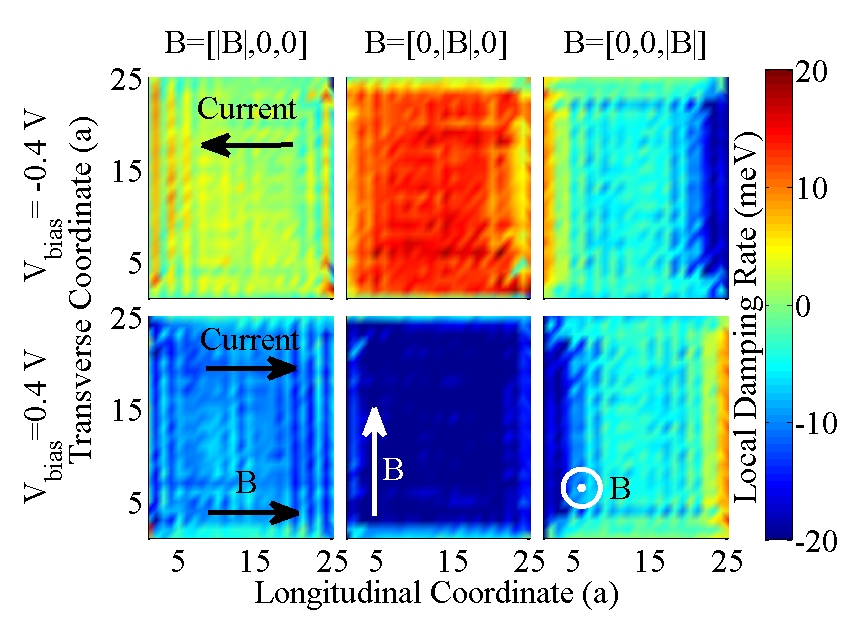}
					\caption{(Color online) Spatial dependence of the local damping rate for the spin-$1/2$ local moments of a FM island under different bias voltages ($\pm 0.4V$) and magnetization directions. For the parameters we chose the size of the FM island to be $25\times25 a^2$, the effective magnetic field, $|B|=20 meV$, the broadening, $\eta$ = 0, and $k_BT=10$ meV. }
					\label{fig:fig7}
				\end{figure}
				Fig.~\ref{fig:fig7} shows the spatial dependence of the spin-$\frac{1}{2}$ local moment switching rate for a  FM/Rashba bilayer (Fig. 1) for two bias values ($V_{bias}=\pm0.4V$) and for an in-plane effective magnetic field (a) along and (b) normal to the transport direction, and (c) an out-of-plane magnetic field. The size of the FM island is  $25a\times25a$, where $a$ is the lattice constant.
 Negative local moment switching rate (blue) denotes that, once excited, the local moment relaxes to its ground state pointing along the direction of the effective magnetic field; however positive local damping rate (red) denotes that the local moments remain in the excited state during the bias pulse duration. Therefore, the damping rate of the local moments under bias voltage can be either enhanced or reduced and even change sign depending on the sign of the bias voltage and the direction of the magnetization. We find that the bias-induced change of the damping rate is highest when the FM magnetization is in-plane and normal to the transport directions similar to the single domain case. Furthermore, the voltage-induced damping rate is peaked close to either the left or right edge of the FM (where the reservoirs are attached) depending on the sign of the bias. Note that there is also a finite voltage-induced damping rate when the magnetization is in-plane and and along the transport direction ($x$) or out-of-the-plane ($z$). 	

				\begin{figure}
				\includegraphics[scale=0.4,angle=0]{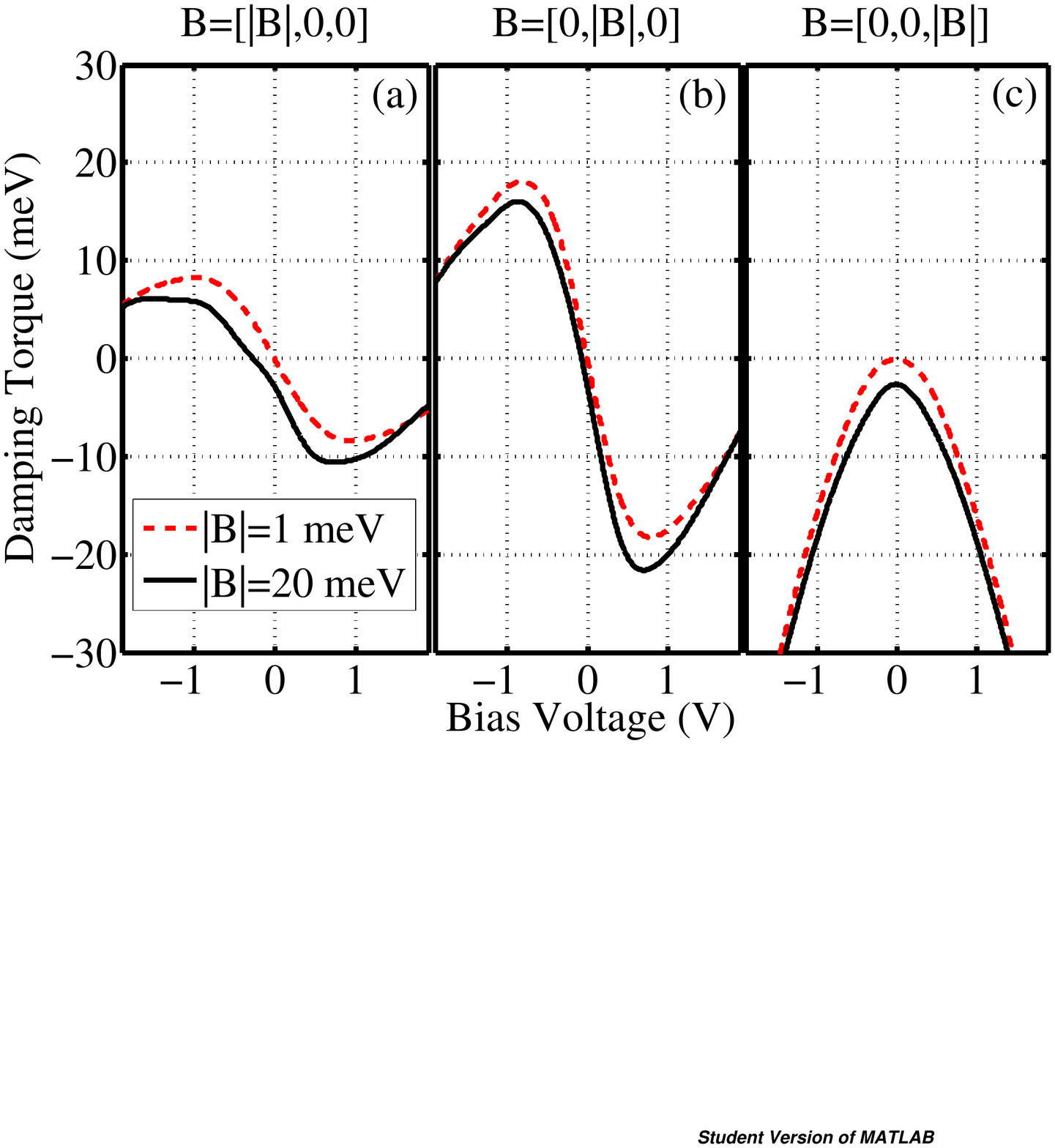}
					\caption{(Color online) Bias dependence of the average (over all sites) damping rate of the FM island for in-plane
effective magnetic field (or equilibrium magnetization) (a) along and (b) normal to the transport direction and (c) out-of-plane magnetic field
for two values of $|B|$. }
					\label{fig:fig8}
				\end{figure}
		Fig.~\ref{fig:fig8} shows the bias dependence of the average (over all sites) damping rate for in- (a and b) and out-of-plane (c) directions of the effective magnetic field (direction of the equilibrium magnetization) and for two values of $|B|$.
This quantity describes the damping rate of the amplitude of the magnetic order parameter. For an in-plane magnetization and normal to the transport direction (Fig.~\ref{fig:fig8}) the bias behavior of the damping rate is linear and finite in contrast to the single domain [Fig.~\ref{fig:fig3}(a)] where the damping rate was found to have a negligible response under bias. On the other hand,
the bias behavior of the current induced damping rate shows similar behavior to the single domain case when the equilibrium magnetization direction is in-plane and normal
to the transport direction (Fig.~\ref{fig:fig8}(b)). For an out-of-plane effective magnetic field [Fig.~\ref{fig:fig8}(c)] the damping
torque has an even dependence on the voltage bias.

%%%%%%%%%%%%%%%%%%%%%%%%%%%%%%%%%%%%%%%%%%%%%%%%%%%%%%%%%%%%%%%%%%%%%%%%%%%%%%%%%%%%%%%%%%%%%%%%%%%%%%%%%%%%%%%%%%%%%%%%%%%%
				
\begin{figure}
				\includegraphics[scale=0.4,angle=0]{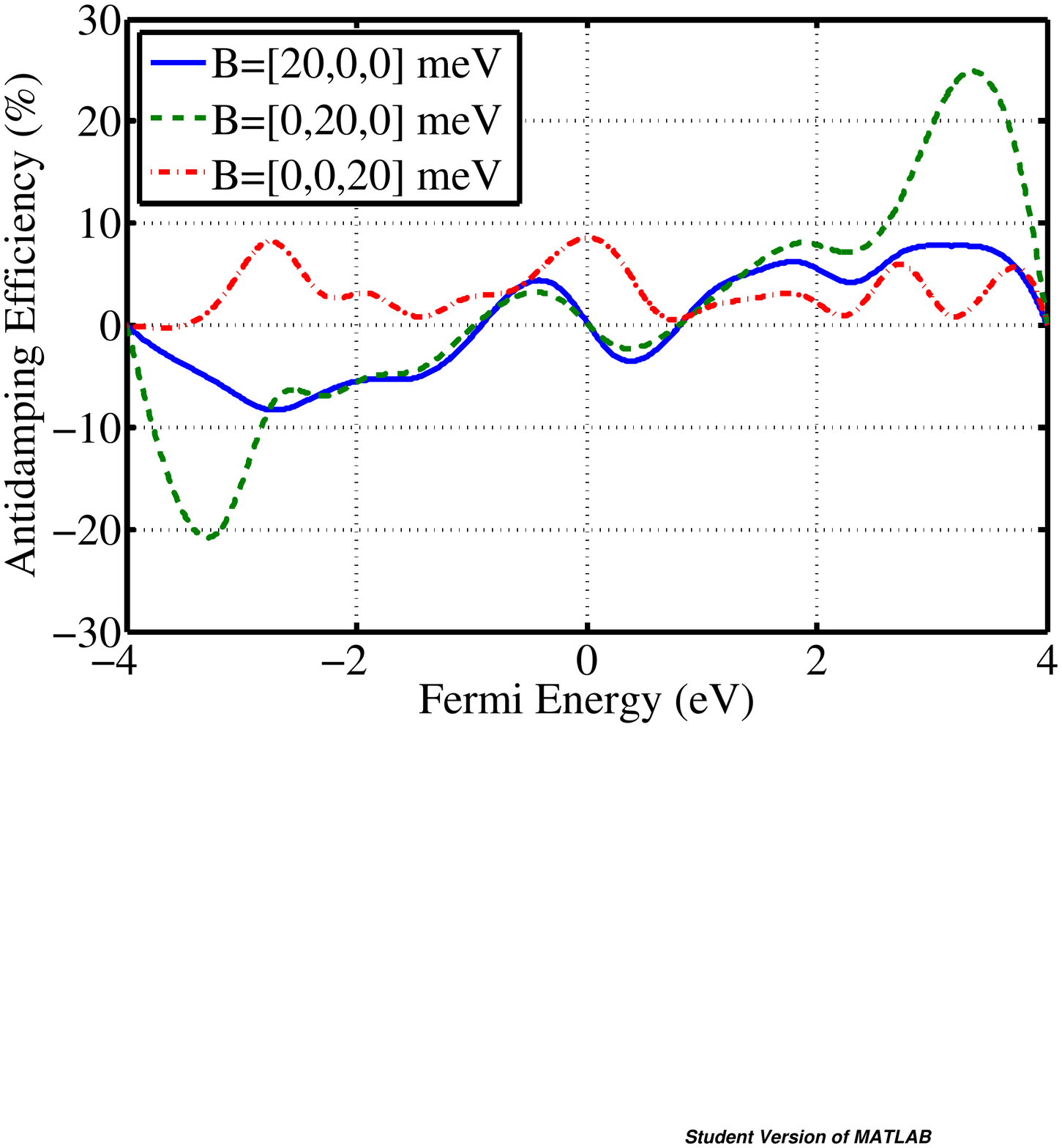}
					\caption{(Color online) Bias-induced local anti-damping efficiency due to local spin-flip, $\Theta=|B|(\mathcal{T}(V_{bias})-\mathcal{T}(0))/eV_{bias}\mathcal{T}(0)$, versus Fermi energy for different equilibrium magnetization orientations. For the calculation we chose $V_{bias}=0.2$ V and the rest of the Hamiltonian parameters are the same as in Fig.~\ref{fig:fig7}.}
					\label{fig:fig9}
				\end{figure}

In order to quantify the efficiency of the voltage induced excitations of the local moments, we calculate the relative change of the average of the damping rate in the presence of a bias voltage and present the result versus the Fermi energy for different orientations of the magnetization in Fig~\ref{fig:fig9}. We find that the efficiency is maximum for an in-plane equilibrium
magnetization normal to the transport direction and it exhibits an electron-hole asymmetry. The bias-induced antidamping efficiency due to spin-flip
can reach a peak around 20\% which is much higher than the peak efficiency of about 2\% in the single domain precession mechanism in Fig.~\ref{fig:fig5} for the same system parameters.

Future work will be aimed in determining the switching phase diagram~\cite{MahfouziPRB2016} by calculating the local antidamping and field-like torques self consistently for different FM configurations.

	\section{Concluding remarks}\label{sec:conclusions}
In conclusion, we have developed a formalism to investigate the current-induced damping rate of nanoscale FM/SOC 2D Rashba plane bilayer in the quantum regime within the framework of the Kyldysh Green function method. We considered two different regimes of FM dynamics, namely, the single domain FM and independent local moments regimes. In the first regime we assume the rotation of the FM as the only degree of freedom, while the second regime takes into account only the local spin-flip mechanism and ignores the rotation of the FM.
 When the magnetization (precession axis) is in-plane and normal to the transport direction, similar to the conventional SOT for classical FMs, we show that the bias voltage can change the damping rate of the FM and for large enough voltage it can lead to a sign reversal. In the case of independent spin-$1/2$ local moments we show that the bias-induced damping rate of the local quantum moments can lead to demagnetization of the FM and has strong spatial dependence. Finally, in both regimes we have calculated the bias-induced damping efficiency as a function of the position of the Fermi energy of the 2D Rashba plane.
	
	\appendix
	\section{Derivation of Electronic Density Matrix}\label{AppA}
	Using the Heisenberg equation of motion for the angular momentum operator, $\vec{\boldsymbol{S}}(t)$, and the commutation relations for the angular momentum, we obtain the following Landau-Lifshitz equations of motion,
	
	\begin{subequations}\label{eq:trans}
		\begin{align}
		&\mp i\frac{\partial}{\partial t}{\boldsymbol{S}}^{\pm}(t)=\boldsymbol{h}^z{\boldsymbol{S}}^{\pm}(t)-{\boldsymbol{h}}^{\pm}(t){\boldsymbol{S}}^z(t)\\
		&-i\frac{\partial}{\partial t}{\boldsymbol{S}}^z(t)=\frac{1}{2}\left({\boldsymbol{h}}^+(t){\boldsymbol{S}}^-(t)-{\boldsymbol{h}}^-(t){\boldsymbol{S}}^+(t)\right)\\
		&\vec{\boldsymbol{h}}_{mm'}(t)=\frac{1}{\hbar}\sum_{\bold{r}}  J_{sd}\vec{\bold{s}}^{mm'}_c(\bold{r})+g\mu_B\delta_{mm'}\vec{B}(t),
		\end{align}
	\end{subequations}
	where, \mbox{${\boldsymbol{S}}^{\pm}={\boldsymbol{S}}_x\pm{\boldsymbol{S}}_y$} (\mbox{${\sigma}^{\pm}={\sigma}_x\pm{\sigma}_y$}), is the angular momentum (spin) ladder operators, $\vec{\boldsymbol{s}}^{mm'}_c(\bold{r})=\frac{1}{2}\sum_{\sigma\sigma'} \vec{\sigma}_{\sigma\sigma'} \rho^{mm'}_{\sigma\sigma',\bold{r}\bold{r}}$ is the local spin density of the conduction electrons which is an operator in magnetic configuration space. Here, $\rho$ is the density matrix of the system, and the subscripts, $\bold{r},m,\sigma$ refer to the atomic cite index, magnetic state and spin of the conduction electrons, respectively. In the following we assume a precessing solution for Eq~\eqref{eq:trans}(a) with a fixed cone angle and Larmor frequency $\omega=h^z$. Extending the Hilbert space of the electrons to include the angular momentum degree of freedom we define ${\psi}_{ms'i}(t)=|S,m\rangle\otimes \psi_{s'i}(t)$. The equation of motion for the Green function (GF) is then given
	\begin{align}
	&\left(E-i\eta-\hat{H}(k)+n\omega-\frac{n}{2S}J_{sd}(k)\sigma^z\right)\hat{\boldsymbol{G}}^r_{nm}(E,k)\\
	&-\frac{\sqrt{S(S+1)-n(n+1)}}{2S} J_{sd}(k)\sigma^-\hat{\boldsymbol{G}}^r_{n+1m}(E,k)\nonumber\\
	&-\frac{\sqrt{S(S+1)-n(n-1)}}{2S}J_{sd}(k)\sigma^+\hat{\boldsymbol{G}}^r_{n-1m}(E,k)=\hat{1}\delta_{nm}\nonumber
	\end{align}
	where,   $n=(-S,-S+1,...,S)$ and the gauge transformation \mbox{$\psi_{n\sigma i}(t)\rightarrow e^{in\omega t}\psi_{n\sigma i}(t)$}
has been employed to remove the time dependence. The density matrix of the system is of the form
	\begin{align}
	\hat{\boldsymbol{\rho}}_{nm}=e^{-i(n-m)\omega t}\sum_{p=-S}^S\int \frac{dE}{2\pi}\hat{\boldsymbol{G}}^r_{np}2\eta f_{p\hat{\mu}}\hat{\boldsymbol{G}}^a_{pm}
	\end{align}
	where, $f_{p\hat{\mu}}(E)=f(E-p\omega-\hat{\mu})$ is the equilibrium Fermi distribution function of the electrons. Due to the fact that $p\omega$ are the eigenvalues of \mbox{ $\boldsymbol{H}_M=-g\mu_B \vec{B}\cdot\boldsymbol{S}$}, one can generalize this expression by transforming into a basis where the magnetic energy is not diagonal which in turn leads to Eq~\eqref{eq:GenDensMat} for the density matrix of the conduction electron-local moment entagled system.

	\section{Recursive Relation for GFs}\label{AppB}
	Since in this work we are interested in diagonal blocks of the GFs and in general for FMs at low temperature we have  $S\gg 1$, we need a recursive algorithm to be able to solve the system numerically. The surface Keldysh GFs corresponding to ascending $\hat{g}^{u,r/<}$, and descending  $\hat{g}^{d,r/<}$, recursion scheme read,

	\begin{widetext}
		
	\begin{align}
	&\hat{g}^{u,r}_{n}(E,k)=\frac{1}{E-\omega_n-i\eta_n-\hat{H}(k)-\hat{\Sigma}_n^r(E,k)-\frac{n}{2S}{ J}_{sd}(k)\sigma^z-\frac{(S^{-}_n)^2}{4S^2}{ J}_{sd}(k)\sigma^+\hat{g}^{u,r}_{n-1}(E,k)\sigma^-{ J}_{sd}(k)}\\
	&\hat{\Sigma}^{u,<}_{n}(E,k)=-\sum_{\alpha}\left(2i\eta_n+\hat{\Sigma}^r_{n,\alpha} (E,k)-\hat{\Sigma}^a_{n,\alpha} (E,k)\right)f_{n\alpha}+\frac{(S^{-}_n)^2}{4S^2}{ J}_{sd}\sigma^+\hat{g}^{u,r}_{n-1}\hat{\Sigma}^{u,<}_{n-1}\hat{g}^{u,a}_{n-1}\sigma^-{ J}_{sd}\\
	&\hat{g}^{d,r}_{n}(E,k)=\frac{1}{E-\omega_n-i\eta_n-\hat{\Sigma}_n^r(E,k)-\hat{H}(k)-\frac{n}{2S}{ J}_{sd}(k)\sigma^z-\frac{(S^{+}_n)^2}{4S^2}{ J}_{sd}(k)\sigma^-\hat{g}^{u,r}_{n+1}(E,k)\sigma^+{ J}_{sd}(k)}\\				
	&\hat{\Sigma}^{d,<}_{n}(E,k)=-\sum_{\alpha}\left(2i\eta_n+\hat{\Sigma}^r_{n,\alpha} (E,k)-\hat{\Sigma}^a_{n,\alpha} (E,k)\right)f_{n\alpha}+\frac{(S^{+}_n)^2}{4S^2}{ J}_{sd}\sigma^-\hat{g}^{d,r}_{n+1}\hat{\Sigma}^{d,<}_{n+1}\hat{g}^{d,a}_{n+1}\sigma^+{ J}_{sd}
	\end{align}
\end{widetext}
	where, $\hat{\Sigma}^r_{n}(E,k)=\sum_{\alpha}\hat{\Sigma}^r_{\alpha}(E-\omega_n,k)$ corresponds to the self energy of the leads, $\alpha=L,R$ refers to the left and right leads in the two terminal device in Fig. \ref{fig:fig3} and $S^{\pm}_m=\sqrt{S(S+1)-m(m\pm1)}$. Using the surface GFs we can calculate the GFs as follows,
	
	\begin{widetext}
	
	\begin{align}
	\hat{\boldsymbol{G}}^{r}_{n,m}(E,k)&=\frac{1}{E-\omega_n-i\eta_n-\hat{H}(k)-\hat{\Sigma}_n^r-\frac{n}{2S}{ J}_{sd}(k)\sigma^z-\hat{\Sigma}^{r,u}_n-\hat{\Sigma}^{r,d}_n},\ \ \ n=m\\
	&=\frac{S^{+}_n}{2S} \hat{g}^{u,r}_{n}(E,k){ J}_{sd}(k)\sigma^-\hat{\boldsymbol{G}}^{r}_{n+1,m}(E,k),\ \ \ \ n\neq m\\
	&=\frac{S^{-}_n}{2S} \hat{g}^{d,r}_{n}(E,k){ J}_{sd}(k)\sigma^+\hat{\boldsymbol{G}}^{r}_{n-1,m}(E,k),\ \ \ n\neq m
	\end{align}
	
	where the ascending and descending self energies are given by,
	\begin{align}
	\hat{\Sigma}^{r,u}_n=\frac{(S^{-}_n)^2}{4S^2}{ J}_{sd}(k)\sigma^+\hat{g}^{u,r}_{n-1}(E,k)\sigma^-{ J}_{sd}(k)\\
	\hat{\Sigma}^{r,d}_n=\frac{(S^{+}_n)^2}{4S^2}{ J}_{sd}(k)\sigma^-\hat{g}^{d,r}_{n+1}(E,k)\sigma^+{ J}_{sd}(k)
	\end{align}
%	Similarly for the density matrix we get,
%	\begin{align}
%	\hat{\boldsymbol{\rho}}_{nn+1}(t)=e^{i\omega t}\int \frac{dE}{\pi}\left(\hat{\boldsymbol{G}}^r_{nn}(E)\hat{g}^{u,<}_{n}(E,k)\hat{\boldsymbol{G}}^a_{nn+1}(E)+\hat{\boldsymbol{G}}^r_{nn+1}(E)\hat{g}^{d,<}_{n+1}(E,k)\hat{\boldsymbol{G}}^a_{n+1n+1}(E)\right)
%	\end{align}

	The average rate of angular momentum loss/gain can be obtained from the real part of the loss of angular momentum in one period of precession,
	\begin{align}
	\mathcal{T}'_n=\frac{1}{2}(\mathcal{T}'^-_{n}-\mathcal{T}'^+_{n})=\frac{1}{2}\Im\left(\sum_{k} Tr[\frac{S^{-}_n}{2S}{\sigma}^{+}{ J}_{sd}(k)\hat{\boldsymbol{\rho}}_{nn+1}(k)-\frac{S^{+}_n}{2S}{\sigma}^{-}{ J}_{sd}(k)\hat{\boldsymbol{\rho}}_{nn-1}(k)]\right)
	\end{align}
	which can be interpreted as the current flowing across the layer $n$.
	\begin{align}
	\mathcal{T}'^{-/+}_n= \sum_k\int \frac{dE}{2 \pi i}\,  \mathrm{Tr}\, \left\{ \left[\hat{\Sigma}^{d/u,r}_{n}(E,k)-\hat{\Sigma}^{d/u,a}_{n}(E,k)\right] \hat{\boldsymbol{G}}^{<}_{nn}(E,k) + \hat{\Sigma}^{d/u,<}_{n}(E)\left[ \hat{\bold{G}}^r_{nn}(E,k)-\hat{\bold{G}}_{nn}^a(E,k) \right] \right \},
	\end{align}
	
\end{widetext}

	\begin{acknowledgments}
	The work at CSUN is supported by NSF-Partnership in Research and Education in Materials (PREM) Grant DMR-1205734,
    NSF Grant No. ERC-Translational Applications of Nanoscale Multiferroic Systems (TANMS)-1160504, and US Army of Defense Grant No. W911NF-16-1-0487.
	\end{acknowledgments}

	%BibTeX
	%Windows:
	%\bibliographystyle{D:/PHYSICS/TEX/BIBTEX/prsty}
	%\bibliography{D:/PHYSICS/TEX/BIBTEX/qttg}
	
	%Linux:
	%\bibliographystyle{apsrev}
	%\bibliography{$HOME/TEX/BIBTEX/qttg}
	
\end{document}